\documentclass[12pt,oneside,reqno]{amsart}

\theoremstyle{definition}

\theoremstyle{remark}

\numberwithin{equation}{section}

\usepackage{amssymb}

\begin{document}

\title{BBGKY equations, self-diffusion and 1/f noise \\
in a slightly nonideal gas}

\author{Yuriy\, E.\, Kuzovlev}
\address{Donetsk Institute for Physics and Technology of NASU,
ul.\,R.\,Luxemburg 72, Donetsk 83114, Ukraine}
\email{kuzovlev@kinetic.ac.donetsk.ua}




\begin{abstract}
The hypothesis of ``molecular chaos'' is shown to fail when applied
to spatially inhomogeneous evolution of a low-density gas, because
this hypothesis is incompatible with reduction of interactions of gas
particles to ``collisions''. The failure of molecular chaos means
existence of statistical correlations between colliding and closely
spaced particles in configuration space. If this fact is taken into
account, then in the collisional approximation (in the kinetic stage
of gas evolution) in the limit of infinitely small gas parameter the
Bogolyubov-Born-Green-Kirkwood-Yvon (BBGKY) hierarchy of equations
yields an autonomous system of kinetic equations for the
many-particle distribution functions of closely spaced particles.
This system of equations can produce the Boltzmann equation only in
the homogeneous case. It is used to analyze statistical properties of
Brownian motion of a test gas particle. The analysis shows that there
exist fluctuations with a 1/f spectrum in the diffusivity and
mobility of any particle. The physical cause of these fluctuations is
randomness of distribution of particles' encounters over the impact
parameter values and, consequently, randomness of the rate and
efficiency of collisions.

In essence, this is {\bf reprint} of the like author's paper
published in Russian in [\,Zh.\, Eksp.\, Teor.\, Fiz.\, {\bf 94}
(12), 140-156 (Dec. 1988)] and translated into English in [\,Sov.\,
Phys.\, JETP\, {\bf 67} (12), 2469-2477 (Dec. 1988)] twenty years ago
but seemingly still unknown to those to whom it might be very useful.
The footnotes contain presently added comments.
\end{abstract}


\maketitle

\baselineskip 24 pt

\markboth{}{}

\section{Introduction}
The today's kinetic theory of weakly nonideal gases as before rests
on the antiquated hypothesis of ``molecular chaos'' which asserts
that the particles entering a collision are statistically independent
and which makes it possible to reduce the exact
Bogolyubov-Born-Green-Kirkwood-Yvon (BBGKY) equations to the
classical model Boltzmann equation \cite{bog,re,ll2}. Meanwhile, the
molecular chaos never has been proved and, in fact, can be justified
only for the special case of spatially homogeneous gas evolution
\cite{re,ll2,kr,kac,bal}. As for the general case, any reasonings in
favor of the molecular chaos involve other arbitrary assumptions. For
example, it is sometimes identified with those fact that in a
sufficiently low-density gas the colliding particles do not have an
intersecting dynamic past. However, as was emphasized many years ago
\cite{kr}, generally the absence of dynamic correlations by no means
implies the absence of statistical correlations from the standpoint
of the probability laws which pertain to an ensemble of dynamic
systems. It is also incorrect to identify molecular chaos with the
decoupling of statistical correlations for infinitely far spaced
particles, since in actuality the particles arrive at a collision not
from infinity but from a distance which is only on the order or less
than the mean free path $\,\lambda_0\,$ (moreover, in practice,
molecular chaos is being postulated even for distances on the order
of the interaction radius $\,r_0\ll\lambda_0\,$)\,\footnote{\,Another
example of the arbitrary assumptions is replacement of the BBGKY
hierarchy by so called ``hard sphere BBGKY hierarchy'' which has no
substantiation but was exploited in the Lanford's attempt to
substantiate the Boltzmann equation (although for absurdly small
evolution time only) [\,O.E.Lanford, ``Time evolution of large
classical systems'', in\, ``Dynamical systems, theory and
applications'', ed. J.Moser, Lectures Notes in Physics, vol.38,
1975;\, ``On a derivation of Boltzmann equation'', in\, ``Nonlinear
phenomena. 1. The Boltzmann equation'', eds. J.L.Lebowitz,
E.W.Montroll, N.-H., Amsterdam, 1983;\, H.Spohn,``Theory of
fluctuations and Boltzmann equation'',\, ibid.\,]. In fact, the
``hard sphere BBGKY hierarchy'' is a ``hand-made'' probabilistic
model which does not follow from the Liouville equation even in the
limit of infinitely hard repulsion. The present paper helps to
comprehend why it is so\, (for detail, see [Yu.\,Kuzovlev,\, ``On
Brownian motion in ideal gas and related principles'',\, arXiv:\,
0806.4157\,] and footnotes below). \label{fn1}}\,.

On the other hand, it is not difficult to indicate why the molecular
chaos can fail in inhomogeneous non-equilibrium situations. Notice
that, first, in spatially inhomogeneous gas configurational (spatial)
dependencies of distribution functions (DF) carry statistical
information not only about the instantaneous coordinates of gas
particles but also, indirectly, about their past diffusive
displacements, or ``Brownian paths'' (since the non-homogeneity
constitutes a natural reference scale for the displacements). Second,
the displacement of each particle is closely correlated with
fluctuations in the rate of collisions of this particle and
consequently is correlated, to the extent of duration of these
fluctuations, with next collisions. Therefore the pair (two-particle)
DF for pre-colliding particles (i.e. particles going into mutual
collision) actually represents a conditional probability distribution
under the conditions that a new collision realizes and it takes place
in a given small space-time region. Even because of this circumstance
alone, generally the mentioned DF can not be factored into the
product of one-particle DF which would furnish only unconditional
information about the coordinates and displacements.

The question thus is the extent to which the fluctuations in the
collision rate of an arbitrary gas particle are ``long-living''. A
natural source of these fluctuations is the randomness of the impact
parameter in an encounter of two particles (we will be using the word
``encounter'' to mean both a factual interaction of particles and
their mere passing at a relative distance comparable with the
interaction radius $\,r_0\,$). It is quite obvious that this source
does not reduce completely to the ordinary local gas density
fluctuations. Being dependent on a random distribution of the
particles' encounters over values of the impact parameter, the actual
rate of collisions of any given particle can randomly vary from one
time interval to another. But, at the same time, a thermodynamic
state of the gas is indifferent to these variations, since anyway
they do not interfere with a local thermalization of the gas during
time on order of the mean free path time $\,\tau_0\,$. Consequently,
there are no relaxation mechanisms which would tend to establish some
certain distribution (histogram) of the impact parameter values, and
in this sense the relaxation time (or smoothing time) of this
distribution is infinitely long. Such the reasonings show that
fluctuations in the collision rate (and thus in the gas kinetic
coefficients associated with this rate) are of a long-living
``flicker'' nature \cite{bk1,bk2,pr195,bkc,bk2}.

To deal with these fluctuations we must abandon {\it a priori}
molecular chaos, i.e. treat the pair DF, for particles which are
encountering one another (in the sense explained above), as an
autonomous statistical characteristics of gas evolution, which
represents a local ensemble-average rate of encounters and thus rate
of collisions proper. According to the BBGKY equations, the evolution
of this DF is coupled with evolution of the higher-order DF for
``clusters'' of $\,n>2\,$ relatively close encountering particles.
One might suggest that, taken together, they represent temporal
statistics of impact parameters of particles'encounters and thus
statistics of the rate of collisions.

In Sec.2 we will show that in the framework of the coarsened
``collisional'' (kinetic) description of the particles' interaction
the BBGKY hierarchy generates a separate system of evolution
equations for just mentioned special DF of particles under mutual
encounters. At that, the structure of these equations is such that in
spatially inhomogeneous case it forbids Boltzmann's molecular chaos.
The only possibility in general is a weakened version of the
molecular chaos hypothesis (examined in Sec.3) which incorporates the
inter-particle statistical correlations in configuration space.

Such a weakened hypothesis, however, is sufficient for deriving a
closed (although infinite) system of kinetic equations. As is shown
in Sec.4 by the example of self-diffusion, these equations predict
flicker (1/f) fluctuations in the transport coefficients of a gas.
This result, discussed in Sec.5, supports the fundamental conception
of 1/f noise which was first proposed in \cite{bk1,bk3}.

\section{Collisional approximation}
Since the BBGKY hierarchy can not be solved rigorously, we naturally
appeal to the idea, which was suggested by Bogolyubov in \cite{bog},
about possibility of asymptotic separation of ``collisional'' and
``kinetic'' space-time scales in the low-density limit\,
($\,\lambda_0=\,$const\,, $\,\mu\equiv r_0/\lambda_0\sim \rho r_0^3
\rightarrow 0\,$, where $\,\rho\,$ is mean gas density). In other
words, following \cite{bog}, let us assume that at a sufficiently
late stage of the gas evolution all many-particle DF $\,F_n\,$
possess, along with $\,F_1\,$, only slow time dependence
characterized by ``kinetic'' time scales $\,\gtrsim \tau_0
=\lambda_0/v_0\,$ ($\,v_0=\sqrt{T/m}\,$ is thermal velocity). In
order to implement this idea into practice we have to specify the
approximate asymptotic form in which we are seeking a solution of the
BBGKY equations. For this purpose, Bogolyubov in \cite{bog}
introduced the assumption that all of the DF $\,F_n\,$ are local, in
respect to time, functionals of the one-particle DF $\,F_1\,$. That
assumption makes a use of the molecular chaos hypothesis unavoidable,
although the BBGKY equations by themselves in no way impose this
hypothesis. The ``slowness'' of $\,F_n\,$, however, gives a longer
list of possibilities. We will accordingly discuss a less rigid
formulation of the separation of scales.

To consider DF for closely spaced particles, it is natural to express
the inter-particle distances $\,q_{ij}=q_{i}- q_{j}\,$ ($\, q_{i}\,$
are coordinates) in units of $\,r_0\,$, while the position of the
center of mass of a cluster as a whole, $\, q^{(n)}=(1/n)\sum_{j=1}^n
q_{j}\,$, in units of $\,\lambda_0\,$. Putting the particle
velocities in dimensionless form by dividing by thermal velocity
scale $\,v_0\,$, and putting the time in dimensionless form by the
mean free time $\,\tau_0\,$ (in accordance with the presumed slowness
of changes of DF), we can write the following expression for the
volume-normalized DF:
\begin{equation}
F_n(t\,,q^{(n)},q_{ij}\,,v_{j})\,=
\,v_0^{-\,3n}\,\widetilde{F}_n\left (\frac {t}{\tau_0}\,, \frac {
q^{(n)}}{\lambda_0}\,,\frac { q_{ij}}{r_0}\,, \frac
{v_{j}}{v_0}\right )\,\,. \label{df}
\end{equation}
The separation of scales presupposes that in a certain asymptotic
sense the ``reduced'' DF of close-lying particles,
$\,\widetilde{F}_n\,$, do not depend on the gas density $\,rho\,$,
i.e. do not contain the gas parameter $\,\mu =\rho r_0^3\,$ as a
special independent argument. Let us examine the consequences of this
- still preliminary - suggestion by substituting (\ref{df}) into the
BBGKY equations.

To exhibit the scale $\,r_0\,$ explicitly, it is convenient to
specify the inter-particle interaction force to be $\,(T/r_0)\,
f(q_{ij}/r_0)\,$. At that, we can always choose $\,r_0\,$ and
$\,\lambda_0\,$ in such a way that the relation $\,\lambda_0 =1/\rho
r_0^2\,$ holds. Let us introduce the designations $\,z_n
=q^{(n)}/\lambda_0\,$ and $\,x_{ij}=q_{ij}/r_0\,$, while retaining
the old notations $\,t\,$ and $\,v_j\,$ for new dimensionless time
$\,t/\tau_0\,$ and velocities $\,v_j/v_0\,$. The BBGKY equations can
then be easy put in the following form:
\begin{equation}
\left (\frac {\partial }{\partial t}\,+\,u_n\,\frac {\partial
}{\partial z_n}\,+\frac {1}{\mu}\,L_n^{\prime}\right
)\widetilde{F}_n\, =\,\sum_{j\,=1}^n\,\frac {\partial }{\partial
v_j}\int\! f(x_{n+\!1\,j})\,\widetilde{F}_{n+\!1}\,\,dx_{n+\!1\,j}\,
\,dv_{n+\!1}\, \equiv\,J_n\,\,,\,\label{fn}
\end{equation}
where operator $\,L_n^{\prime}\,$ (which acts on the functional
dependence on $\,x_{ij}\,$ and $\,v_j\,$ only) is the Liouville
operator of the relative motion and interaction of $\,n\,$ particles,
which results from the complete $\,n$-particle Liouville operator by
eliminating the center-of-mass motion, and
\[
u_n\,=\,\frac 1n \,\sum_{j\,=1}^n v_j\,\,
\]
is the center-of-mass velocity.

It is seen from equations (\ref{fn}) that formally strict
independence of $\,\widetilde{F}_n\,$ on $\,\mu\,$ would imply a
supplement to the equations in the form of the requirement
\begin{equation}
L_n^{\prime}\,\widetilde{F}_n\, =\,0\,\,,\,\label{req}
\end{equation}
which just excludes contributions of ``fast'' relative interactive
motion of particles. The physical meaning of this requirement is easy
understandable. It asserts that different dynamical states, which
realize in the course of an encounter of $\,n\,$ particles on the
same phase trajectory in $\,n$-particle phase space, have the same
probability in the statistical ensemble under consideration . In
other words, different dynamic stages of the same collision (in
particular, {\it in\,}- and {\it out\,}-\,states) are represented by
the statistical ensemble with equal weights.

In essence, this statistical property is obligatory attribute of such
a gas evolution which allows coarsened description in terms of
collisions (when details of geometry and time history of the
interaction process are replaced by indication of only input and
output of a momentary ``collision''). Therefore we can expect that
exact solution of the BBGKY hierarchy asymptotically satisfies the
requirement (\ref{req}), and thus (\ref{req}) should be thought of as
more adequate basis of the collisional approximation than that
proposed in \cite{bog} (indeed, the alternative to (\ref{req}) would
be unfitness of the concept of collision at all, which would
contradict the elementary physical logics).

At the same time, undoubtedly, the equality (\ref{req}) never holds
in literal rigorous sense, at least because the presupposed scale
separation concerns only not too long inter-particle distances, at
any case $\,|q_{ij}|\ll\lambda_0\,$ ($\,|x_{ij}|\ll\mu^{-1}\,$). We
will thus move on to a more correct treatment of the separation. For
this purpose it is quite sufficient to understand (\ref{req}) as the
condition that the quantity
$\,\mu^{-1}\,L_n^{\prime}\,\widetilde{F}_n\,$ (or
$\,L_n^{\prime}\,F_n\,$ in the original dimensional form) is small in
comparison with the other terms of the $\,n$-th BBGKY equation.
Furthermore, it is sufficient if (\ref{req}) holds only on the
average over some region in the space $\,q_{ij}\,$ with a linear size
$\,a\,$ much larger than $\,r_0\,$ but much smaller than
$\,\lambda_0\,$. A natural (and unambiguous in order of magnitude)
choice for  $\,a\,$ if the average distance between neighboring
particles: $\,a=\rho^{-1/3}\,$ (in the dimensionless form,
$\,a/r_0=\mu^{-1/3}\,$).

In the limit $\,\mu\rightarrow 0\,$ this region (the ``collision
volume'') becomes infinitely large on the scale of $\,r_0\,$ but it
shrinks to a point at the scale of $\,\lambda_0\,$. Then one can
neglect the vanishingly small ($\,\lesssim a/\lambda_0\sim
\mu^{2/3}\,$) difference between the centers of mass of the
configurations on the left and right sides of (\ref{fn}), and replace
the chain of variables $\,z_n\,$ by the single common variable
$\,z\,$\,: the coordinate of a physically small collision volume. The
belonging to the same such volume will be taken below as the
criterium of closeness of particles.

Let us denote the mentioned averaging operation by the overline, and
the result of the averaging of $\,\widetilde{F}_n\,$ by
$\,A_n=\overline{\widetilde{F}_n}\,$. By virtue of this definition of
DF $\,A_n\,$, any of $\,A_n=A_n(t,z,v_1,...,v_n)\,$ depend only on
$\,t\,$, $\,z\,$ and the velocities $\,v_j\,$ and characterizes a
local mean (ensemble-average) density of the number of $\,n$-particle
encounters. At that, according to the aforesaid, now instead of
(\ref{fn}) we have
\begin{equation}
\overline{\mu^{-1}\,L_n^{\prime}\,\widetilde{F}_n}\,
=\,0\,\,,\,\label{req1}
\end{equation}
(or $\,\overline{L_n^{\prime}\,F_n}\,$ in the dimensional form). Due
to this equality the equations (\ref{fn}) turn into
\begin{equation}
\left (\frac {\partial }{\partial t}\,+\,u_n\,\nabla \right )A_n\,
=\,\overline{J_n}\,\,,\,\label{an}
\end{equation}
where $\,\nabla =\partial/\partial z\,$, and (as in (\ref{req1})) the
limit $\,\mu\rightarrow 0\,$ is taken in mind.

It is thus clear that in general inhomogeneous case, when $\,\nabla
A_n \neq 0\,$, a solution of equations (\ref{an}) can not be written
as the product of one-particle DF:
\[
A_n(t,z,v_1,...,v_n)\,\neq\,\prod_{j=1}^n A_1(t,z,v_j)\,\,,
\]
since the inertial terms $\,\,u_n\,\nabla A_n\,$ constantly generate
statistical correlations between close-lying particles, due to their
joint drift (together with their collision volume) relative to the
inhomogeneity. As the consequence, the circumstance that particles
belong to the same encounter or collision event already establishes
statistical correlations between them. At that, of course, from the
probabilistic point of view, there is no principal difference between
encounters and collisions proper.

We thus arrive at the conclusion that in non-homogeneous situations
the collisional approximation - by virtue of its very nature -
contradicts the hypothesis of molecular chaos,\, since in the
language of collisions the relative motion of colliding particles
becomes an inner constituent part of the collision as a whole and
therefore automatically excluded from the equations for DF what
characterize number density of collisions. As a result, even in the
hydrodynamic stage, the evolution of gas is described by the infinite
system of equations (\ref{an}).

But, on the other hand, the collisional character of gas evolution
does not prevent factorization of the $\,n$-particle DF
$\,F_n(t,q_1,...,q_n,v_1,...,v_n)\,$ if the particles are
sufficiently far apart from each other, i.e. are not close in the
above-defined sense. Indeed, if in the limit
$\,r_0/\lambda_0\rightarrow 0\,$ ($\,\lambda_0=\,$const) one keeps
the inter-particle distances $\,q_{ij}\,$ fixed in units of
$\,\lambda_0\,$, instead of $\,r_0\,$, then the BBGKY equations
reduce to such the equations for $\,F_n\,$ which have the factored
solution $\,F_n=\prod_j F_1(t,q_j\,,v_j)\,$. However, in the (right
side of) equation for $\,F_1\,$ we see the pair DF taken at quite
different type of limit, when $\,q_{12}\,$ is fixed in units of
$\,r_0\,$, i.e. $\,F_2|_{\,q_2\,=\,q_1}\,$\,, which leads to the
hierarchy (\ref{an})\,\footnote{\, Notice that our reasonings nowhere
appeal to details of the interaction potential (presuming only that
it is short-range enough). Therefore nothing prevents us to extend
our conclusions to the limit of the hard-sphere interaction.
Moreover, there is no alternative, since only ansatz like (\ref{req})
ensures that probabilities are conserved during collisions.
 \label{fn2}}\,.

The switch to the common spatial variable $\,z\,$ in (\ref{an}), of
course, presumes that $\,|q^{(n)}-q^{(n+1)}|\ll l\,$ and $\,na^3\ll
l^3\,$, where $\,l\,$ is characteristic scale of the non-homogeneity
and $\,na^3\,$ is characteristic volume of a cluster of $\,n\,$ close
particles. In the limit under consideration, both these requirements
are satisfied by an infinite margin if $\,l\gtrsim \lambda_0\,$
(since then $\,na^3/l^3\sim n\mu^2\rightarrow 0\,$) and thus do not
restrict the number of DF $\,A_n\,$ which are ``tied'' to a given
coordinate $\,z\,$.

\section{Weakened molecular chaos}
The requirements (\ref{req1}) are main tools of construction of a
collisional approximation: due to them the right sides of (\ref{fn})
and (\ref{an}) can be reduced to the collisional form. In particular,
with $\,n=2\,$, Eq.\ref{req1} becomes
\[
\overline{\mu^{-1}\,L_n^{\prime}\,\widetilde{F}_n}\,
=\,a^{-3}\mu^{-1}\int_{\,|q_{21}|<\,a}
L_n^{\prime}\,\widetilde{F}_n\,dq_{21}\,=\,\int_{\,|x_{21}|<\,a/r_0}
L_n^{\prime}\,\widetilde{F}_n\,dx_{21}\,=0\,\,
\]
or, after we take the limit $\,\mu\rightarrow 0\,$,
\[
\int_{\,|x_{21}|<\,\infty} \left[(v_2-v_1)\,\frac {\partial}{\partial
x_{21}}\, +\,f(x_{21})\left(\frac {\partial}{\partial v_2} -\frac
{\partial}{\partial v_1} \right )\right ]
\widetilde{F}_n\,dx_{21}\,=0\,\,.
\]
With the help of this equality the right side
$\,\overline{J_1}=J_1\,$ in the first of equations (\ref{fn}) and
(\ref{an}) transforms to integral
\[
\overline{J_1}\,=\,\int dv_{\,2}\,(v_2-v_1)\oint
ds\,\widetilde{F}_2\,
\]
over an infinitely remote surface $\,|x_{21}|=\infty\,$ (with
$\,ds\,$ being its normal vector), so that $\,\overline{J_1}\,$ is
determined by the particle flow into the ``collision volume''
$\,|x_{21}|<\infty\,$ from the surrounding gas.

Depending on the sign of the scalar product $\,(v_2-v_1)\cdot ds\,$,
the DF $\,\widetilde{F}_2\,$ represents either {\it in\,}-\, or\,
{\it out\,}-state of particle 2 with respect to particle 1. Let us
denote by\, $\,A_2^{in}\,$ the values of $\,\lim_{\mu\rightarrow 0}
\widetilde{F}_2\,$ on that part of the boundary surface
$\,|q_{21}|\simeq a\,$ which corresponds to {\it in\,}-states. The
boundary values for the {\it out\,}-states can then be expressed in
terms of the $\,A_2^{in}\,$ with the help of the two-particle
scattering matrix. After that, $\,\overline{J_1}\,$ acquires the
standard form of the collision integral:
\[
\left (\frac {\partial }{\partial t}\,+\,u_1\,\nabla \right )A_1\,
=\,\int dv_{\,2}\, \widehat{S}_{12}\,A_2^{in}\,\,.
\]
Here and below $\,\widehat{S}_{ij}\,$ is the ordinary ``Boltzmann
collision operator'' for the collision of particles $\,i\,$ and
$\,j\,$. The action of this operator is defined by \cite{bog,re,ll2}
\[
\widehat{S}_{ij}\,\psi(v_i\,v_j)\,=\,|v_i-v_j|\, \int d^2b\,\,
[\,\psi(v_i^{\prime}\,v_j^{\prime})-\psi(v_i\,v_j)\,]\,\,,
\]
where $\,b\,$ is the two-dimensional impact parameter vector, and
$\,v_i^{\prime}\,$ and $\,v_j^{\prime}\,$ are the initial velocities
which correspond to the final velocities $\,v_i\,$ and $\,v_j\,$.

Analogously, we can use (\ref{req1}) with $\,n>2\,$ to perform
similar transformations of the integrals $\,\overline{J_n}\,$. In the
limit $\,\mu\rightarrow 0\,$ the functions $\,A_n\,$ are determined
by the average of $\,\widetilde{F}_n\,$ over an infinite
($\,3(n-1)$-dimensional) region of dimensionless inter-particle
distances. Therefore the result of this averaging represents only
such (pre-\, or post-collisional) configurations of $\,n\,$ particles
where none of them are just now in a collision. Correspondingly, only
two-particle collisions (between some of $\,n\,$ lefthanded particles
in (\ref{an}) and ``external'' righthanded $\,(n+1)$-st particle from
the rest of the gas) contribute to $\,\overline{J_n}\,$. We thus find
what could have predicted earlier:
\begin{equation}
\left (\frac {\partial }{\partial t}\,+\,u_n\,\nabla \right )A_n\,
=\,\sum_{j\,=1}^n \int dv_{n+1}\,\,
\widehat{S}_{j\,n+\!1}\,A_{\,n+\!1}^{in} \,\,,\,\label{kn}
\end{equation}
where $\,A_{\,n+\!1}^{in} \,$ is the boundary DF (similar to
$\,A_2^{in} \,$) representing configurations with the external
particle which always is in infinitely remote\, {\it in\,}-state in
respect to other $\,n\,$ particles. To underline the particular role
of the external particle, we will distinguish its velocity among
other arguments of $\,A_{\,n+\!1}^{in}\,$ and write
$\,A_{\,n+\!1}^{in}=A_{\,n+\!1}^{in} (t,z,v_1\,,...,v_n|\,v_{n+\!
1})\,$.

In order to transform (\ref{kn}) into a closed system of equations,
we have to relate the right-side boundary DF $\,A_{\,n+\!1}^{in} \,$
to the left-side functions. In this step - after transition to the
collision integrals - we need to invoke the concept of molecular
chaos. Concretely, let us assume that the external $\,(n+1)$-st
particle, due to its just noted specificity, has no velocity
correlations with the other particles:
\[
A_{\,n+\! 1}^{in}(t,z,v_1\,,...,v_n|\,v_{n+\! 1})\,
=\,A_1(t,z,v_{n+\! 1})\,A_{\,n}^{\prime}(t,z,v_1\,,...,v_n) \,\,.
\]
However, this velocity factorization does not mean absolute
statistical independence, since it still allows a spatial
correlation, by virtue of which the function $\,A_{\,n}^{\prime}\,$
may differ from $\,A_{\,n}\,$ (according to the definition,
$\,A_{\,n}^{\prime}\,$ is the conditional $\,n$-particle DF
corresponding to the condition that a collision with an additional
particle takes place).

In the ``pure'' form the correlation of particles in the
configuration space is described by the DF integrated over all
velocities. Since in all the configurations under consideration the
particles are infinitely close together from the standpoint of the
scale $\,\lambda_0\,$, the degree of their spatial correlations in
all these configurations should be the same. This statement is
expressed by the equality
\[
\int A_{\,n+\! 1}^{in}\,\, dv_1\,...\,dv_{n+\! 1}\, =\, \int
A_{\,n+\! 1}\,\, dv_1\,...\,dv_{n+\! 1}\,\,.
\]
In essence, it claims the conservation of the number of particles in
the collision processes (notice that $\,A_{\,n+\! 1}\,$ is indirect
characteristics of intermediate stages of encounters and collisions).
This equality makes it possible to relate $\,A_{\,n+\! 1}^{in}\,$ to
$\,A_{\,n+\! 1}\,$ and thus $\,A_{\,n}^{\prime}\,$ to $\,A_{\,n+\!
1}\,$.  It is easy to see that unambiguously simplest form of the
relationships is
\[
A_{\,n}^{\prime}(t,z,v_1\,,...,v_n)\, =\,\int A_{\,n+\!
1}(t,z,v_1\,,...,v_n\,,v_{n+\! 1})\,dv_{n+\! 1} \,\left(\int
A_1(t,z,v_{1})\,dv_{1} \right)^{-1}
\]
or, equivalently,
\begin{equation}
A_{\,n+\! 1}^{in}(t,z,v_1\,,...,v_n|\,v_{n+\! 1})\,=\, \frac
{A_1(t,z,v_{n+\! 1})}{\int A_1(t,z,v)\,dv}\,\int A_{\,n+\!
1}(t,z,v_1\,,...,v_n\,,v)\,dv \,\,.\label{rel}
\end{equation}
This relationship does not touch on the correlations between members
of the left-side $\,n$-particle cluster.

Expression (\ref{rel}) is a weakened version of the hypothesis of
molecular chaos. It incorporates the spatial statistical correlations
of colliding particles, i.e. it asserts that only their velocities
and momenta are statistically independent, but not their coordinates
(thus, all the $\,n+1\,$ particles may be mutually dependent in the
configuration space).

Along with (\ref{rel}), Eqs.\ref{an} form a closed - we wish to
stress this closure - hierarchy of kinetic equations. In the limit of
spatially homogeneous gas, this hierarchy permits the completely
factored solution, $\,A_{\,n}(t,v_1\,,...,v_n)=\prod_{j=1}^n
A_1(t,v_j)\,$, and becomes equivalent to the Boltzmann equation
\[
\frac {\partial A_1(t,v_1)}{\partial t}\,=\,\int dv_2\,\,
\widehat{S}_{12}\, A_1(t,v_1)\,A_1(t,v_2)\,\,.
\]
Moreover, this equation follows already from the first of of formulas
(\ref{rel}) when we note that the conditions of inter-consistent
normalization of the set of DF,
\[
\Omega^{-1}\int\int F_{n+\! 1}\, dq_{n+\! 1}\,dv_{\,n+\! 1}
\,=\,F_n\,\,
\]
(with $\,\Omega\,$ being total, infinite, volume of the system), can
be reduced in the homogeneous limit to the local form $\,\int F_{n+\!
1}\, dv_{\,n+\! 1} \,=\,F_n\,$ ($\,F_0=1\,$). This implies $\,\int
A_{n+\! 1}\, dv_{\,n+\! 1} \,=\,A_n\,$, and from (\ref{rel}) with
$\,n=1\,$ we have\, $\,A_2^{in}(t,v_1|\,v_2)
=A_1(t,v_1)A_1(t,v_2)\,$.

In general non-homogeneous case, such relations no longer hold, since
the exact global form of the conditions of mutual consistency of DF
cannot be replaced by a spatially local form. The evolution of
one-particle DF is determined by the entire infinite chain of
equations (\ref{an}) and (\ref{rel}), and becomes definitely
non-Marcovian, in contrast with evolution in the Boltzmann model.
Clearly, this then leads to a low-frequency temporal dispersion of
the spatially non-local kinetic transport coefficients of the gas. In
turn, this dispersion may serve as a source of information about the
low-frequency fluctuations of the kinetic coefficients, as we will
see below. Of course, it would be wrong to think about the spatial
non-homogeneity as a cause of these fluctuations (they take place
also in homogenous and equilibrium states). In fact, the
non-homogeneity gives only the means by which they manifest
themselves in the one-time DF $\,F_n\,$, due to dependencies of
$\,F_n\,$ in non-homogeneous non-equilibrium states on the kinetic
coefficients. Next, let us consider a simplest non-homogeneous
problem concerning self-diffusion of gas particles.

\section{1/f noise of self-diffusion}
To analyze self-diffusion we need to eliminate from the kinetic
equations the hydrodynamic modes associated with the five integrals
of motion of the system as whole. This can be done easily by taking
the known formal approach (see e.g. \cite{re}): replacing the
probability distribution of the external particle in the collision
integral by the equilibrium one-particle DF. In our notation,
replacing (\ref{rel}) by
\begin{equation}
A_{\,n+\! 1}^{in}(t,z,v_1\,,...,v_n|\,v_{n+\! 1})\,=\, A_0(t,v_{n+\!
1})\int A_{\,n+\! 1}(t,z,v_1\,,...,v_n\,,v)\,dv \,\,, \label{rel0}
\end{equation}
where $\,A_0(v)=(2\pi)^{-\,3/2}\,\exp{(-\,v^2/2)}\,$ is the
equilibrium Maxwell velocity distribution.

Physically, this replacement describes a situation in which the gas
is in equilibrium state in the macroscopic thermodynamical sense.
There is only a small perturbation of the statistical equilibrium
with regard to a single marked ``test'' particle and its immediate
surroundings. The statistical state of the surroundings will be
described by the set of DF which stand on the left sides of
(\ref{an}) and (\ref{kn}). The rest of the gas serves as the
thermostat. Clearly, if the one-particle DF is assigned to the test
particle, then the higher-order DF represent clusters consisting of
the test particle and $\,n-1\,$ other particles from its
surroundings. Thus we assign the index 1 to our test particle.

Further, we also will make use of the Green-Kubo theorem, according
to which (see e.g. \cite{bal}) the generalized diffusion coefficient
$\,\widehat{D}(\tau,\nabla)\,$, which figures in the general
non-local form of the self-diffusion equation (see e.g. \cite{zub}),
\begin{equation}
\frac {\partial W(t,R)}{\partial t}\,=\, \nabla\! \int_0^t
\widehat{D}(t-\tau,\nabla) \nabla \,\,W(\tau,R)\,d\tau\,\,\,,
\label{de}
\end{equation}
- with $\,\nabla =\partial/\partial R\,$ and $\, W(t,R)\,$ denoting
probability density distribution of coordinate of the diffusing
particle,\, - can be related to the linear response of DF of the test
particle to an infinitely weak (potential) force $\,f_{ex}(q_1)\,$
applied to it (we take in mind, as usually, that the force is
``switched on'' at some time moment, e.g. $\,t=0\,$, before which the
gas was in all respects at the equilibrium). Namely, it is a
straightforward matter to prove that the following relation holds
(see Appendix A):
\begin{equation}
\int_0^{\infty} dt\, e^{-\,p\,t} \int dq_{1}\, e^{-\,ikq_1}\int v_1\,
F_1(t,q_1\,,v_1)\, dv_1\,=\,\frac {D(p\,,ik)\,
\widetilde{f}_{ex}(k)}{[\,p\,+D(p\,,ik)\,k^2\,]\,T} \,\,,\label{gk}
\end{equation}
where $\,\widetilde{f}_{ex}(k)\,$ is the Fourier transform of
$\,f_{ex}(q_1)\,$ and
\[
D(p\,,\nabla)\,=\,\int_0^{\infty} d\tau\,
\widehat{D}(\tau,\nabla)\,\,.
\]

To find the response we have to add terms\, $\,m^{-1}\,f_{ex}(q_1)
\,\partial F_n/\partial v_1\,$\, to the left sides of the BBGKY
equations. In the kinetic equations (\ref{kn}), i.e. in the framework
of the collisional approximation under the low-density limit (and in
the dimensionless notation), these terms look as\, $\,f_{ex}(z)
\,\partial A_n/\partial v_1\,$. The replacement of $\,q_1\,$ by
$\,z\,$ in the argument of $\,f_{ex}\,$ presupposes that a change of
$\,f_{ex}\,$ over length scales $\,\lesssim a\,$ is negligibly small
and that the external force has no substantial effect on the dynamics
of collisions. These assumptions are well legitimate in the limits
$\,a\,f_{ex}/T\rightarrow 0\,$ and $\,\mu\rightarrow 0\,$ if, in
addition, characteristic spatial scale (wave length) of the force,
$\,k\,$, is not too small: $\,|k|^{-1}\gtrsim \lambda_0\,$.

After the relations (\ref{rel0}) are substituted into Eqs.\ref{kn},
the Boltzmann collision integral on the right sides transforms into a
generalized Fokker-Planck operator $\,\Lambda\,$ (or the
``Boltzmann-Lorentz operator'' \cite{re}) defined by
\[
\Lambda_j\,\psi(v_j)\,= \,\int dv_{n+1}\,\, \widehat{S}_{j\,n+\!1}\,
\, \psi(v_j)\,A_0(v_{n+1})\,\,.
\]
As the result, we come to the equations (in the dimensionless form)
\begin{equation}
\left (\frac {\partial }{\partial t}\,+\,\left[\frac
{1}{n}\sum_{j\,=1}^n v_j\right]\frac {\partial}{\partial z}\,+\,
f_{ex}(z)\,\frac {\partial}{\partial v_1} \right )A_n\,
=\,\sum_{j\,=1}^n \Lambda_j \int A_{\,n+\!1}\,\, dv_{n+1}
\,\label{ln}
\end{equation}
with the equilibrium initial conditions
\[
A_n|_{t\,=0}\,=\,A_n^0\, \equiv\,\prod_{j\,=1}^n A_0(v_j)\,\,.
\]
In principle, $\,D\,$ can be found directly from evolution of an
non-equilibrium state in absence of external forces. This approach,
however, is less convenient since it requires special consideration
of initial stage of the evolution before its kinetic stage.

Next, consider the mentioned linear response, setting on\,
$\,A_n=A_n^0+\phi_n\,$ ($\,\phi \rightarrow 0\,$), taking the Fourier
transform in $\,z\,$ and the Laplace transform in $\,t\,$, and
denoting Fourier transforms of $\,\phi_n\,$ by
$\,\widetilde{\phi}_n\,$. Then Eqs.\ref{ln} yield
\begin{equation}
\left (p\,+\,\frac {ik}{n}\sum_{j\,=1}^n v_j \right )
\widetilde{\phi}_n\, =\,\sum_{j\,=1}^n \Lambda_j \int
\widetilde{\phi}_{\,n+\!1}\,\, dv_{n+1}\,
+\,v_1\,\widetilde{f}_{ex}\,p^{-\,1}\,A_n^0 \,\,.\label{fln}
\end{equation}
Since these equations cannot be solved in their general form, we will
simplify the problem. First, we restrict the analysis to the first
two terms in the expansion of $\,D\,$ in the gradient of the force
inhomogeneity:
\[
D(p,ik)\,=\, D_0(p)\,+\, (ik)^2\,D_1(p)\,+\,...\,\,.
\]
Second, we choose the simplified model form of the operator
$\,\Lambda\,$, namely, the Einstein-Fokker-Planck (EFP) operator:
\[
\Lambda_j\,=\, \gamma\,\left(\frac {\partial }{\partial v_j}\,v_j\,
+\,\frac {\partial^2 }{\partial v_j^2}\right )\,\,.
\]
Here we have used that\, $\,\Lambda_j\,A_0(v_j)=0\,$.

In reality, such the operator could not arise from a short-range
interaction potential (since it corresponds to scattering through
only small angles). This circumstance, however, should not corrupt
the qualitative side of our results, because all that is required of
$\,\Lambda\,$ is that it ensures relaxation of velocities to thermal
equilibrium. This choice is convenient in that all the eigenfunctions
of $\,\Lambda\,$ turn to product of $\,A_0\,$ and polynomials. We may
recall that the operator $\,\Lambda\,$ corresponding to the Maxwell
interaction potential has the same property \cite{re} (therefore the
following calculations can be generalized to this case). But the EFP
operator has a further advantage: it makes it possible to exploit the
separation of variables and thus to work with only projections of
vector variables onto the force's wave vector $\,k\,$, i.e. to deal
with formally one-dimensional problem (notice that if $\,f_{ex}\,$ is
a potential force then vector $\,\widetilde{f}_{ex}\parallel k\,$).

Our next step is expansion of the response in the wave vector of the
force-induced inhomogeneity:
\[
\widetilde{\phi}_n\, =\,\sum_{N=\,0}^{\infty} (ik)^N
C_n^{(N)}\,p^{-\,1}\,\widetilde{f}_{ex}\, \,.
\]
Using formula (\ref{gk}) to relate this series to the above expansion
of the diffusion coefficient, we find
\[
D_0\,=\int v_1 C_1^{(0)}\,dv_1\,\,\,, \,\,\,\,\,D_1+D_0^2/p\,= \int
v_1 C_1^{(2)}\,dv_1\,\equiv\,\delta_1\,\,.
\]
Inserting the same series into (\ref{fln}), we come to equations
\begin{eqnarray}
p\,C_n^{(0)}\,=\sum_{j\,=1}^n \Lambda_j \int C_{\,n+\! 1}^{(0)}\,
dv_{\,n+\! 1}\,+\,v_1 A_n^0\,\,\,,\nonumber \\
p\,C_n^{(N)}\,=\sum_{j\,=1}^n \Lambda_j \int C_{\,n+\! 1}^{(N)}\,
dv_{\,n+\! 1}\,-\,C_n^{(N\!-1)}\,\frac {1}{n}\sum_{j\,=1}^n
v_j\,\,\,.\nonumber
\end{eqnarray}
The equations from the first row can be solved easily with our choice
of $\,\Lambda\,$ and yield
\begin{equation}
C_n^{(0)}\,=\frac {v_1\,A_n^0}{p+\gamma }\,\,\,,\,\,
\,\,\,D_0(p)\,=\frac {1}{p+\gamma }\,\,.\label{c0}
\end{equation}
The solution of the second row of equations for $\,C_n^{(1)}\,$
should be sought in the form
\[
C_n^{(1)}\,=\left (\alpha_n\, v_1\sum_{j\,=1}^n v_j\,
+\,\beta_n\right ) A_n^0 D_n\,\,,
\]
where $\,\alpha_n\,$ and $\,\beta_n\,$ are functions of $\,p\,$
alone. For them we find the equations
\[
p\,\beta_n\,=\,2\gamma\alpha_{\,n+\! 1}\,\,\,,\,\,
\,\,\,\,p\,\alpha_n\, =-\,2\gamma\alpha_{\,n+\! 1}\,-1/n\,\,\,,
\]
and then
\[
\alpha_n\,=-\frac 1p \,\sum_{j\,=n}^{\infty} \,\frac 1j \left(-\frac
{2\gamma }{p}\right)^{j-n}\,\,\,, \,\,\,\,\,\alpha_n+\beta_n\,
=-\frac {1}{p\,n}\,\,\,.
\]

Now consider the functions
\[
\delta_n\,=\int v_1 C_n^{(2)}\,dv_1\,...\,dv_n\,\,\,,
\]
the first of which determines $\,D_1\,$ (see above). Multiplying the
chain of equations for $\,C_n^{(2)}\,$ by $\,v_1\,$ and integrating
over all velocities, we find, after some algebra,
\[
p\,\delta_n\,=-\gamma\delta_{\,n+\!
1}\,-n^{-1}\,[\,(n+2)\,\alpha_n\,+\,\beta_n\,]\,D_0\,\,\,.
\]
Substituting the expressions found above for $\,\alpha_n\,$ and
$\,\beta_n\,$ into this equation, we find $\,\delta_1\,$ and thus
$\,D_1\,$ in the form of the repeating sum:
\[
D_1+\frac {D_0^2}{p}\,=\frac {D_0}{p^{\,2}} \sum_{n\,=1}^{\infty}
(-X)^{n-1}\left[\frac {1}{n^2}+ \frac {n+1}{n} \sum_{j\,=n}^{\infty}
\frac 1j \,(-2X)^{j-n}\right]\,\,\,,
\]
where\, $\,X\equiv \gamma/p\,$. Then, with the help of the identity
\[
n^{-1}X^{n-1}\,=\,X^{-1}\int_0^X y^{n-1}\, dy\,\,\,,
\]
we can transform the series over $\,n\,$ to the easy summable form:
\begin{eqnarray}
D_1+\frac {D_0^2}{p}\,=\frac {D_0}{p^{\,2}X}\int_0^X
\sum_{n\,=1}^{\infty} (-y)^{n-1} \left[\frac {1}{n}+ \frac {n+1}{n}
\sum_{j\,=n}^{\infty} (-2y)^{j-n}\right]\,dy\,=\,\nonumber \\
=\frac {D_0}{p\,\gamma} \int_0^{\,\gamma/p} \left\{ \frac
{\ln{(1+y)}}{y} + \frac {1}{1+2y}\left[\,\frac {\ln{(1+y)}}{y} +\frac
{1}{1+y}\,\right] \right\} \,\,\,.\label{sum}
\end{eqnarray}

We have gone into the details of the calculations to demonstrate the
characteristic $\,p\,$ dependence of the response which is rather
general and holds under another choice of the operator $\,\Lambda\,$.
Let us consider this dependence at $\,|p|\ll\gamma\,$ and thus
examine the behavior of the diffusion coefficient at low frequencies.
From (\ref{c0}) and (\ref{sum}) at $\,p/\gamma\rightarrow 0\,$ we
have $\,D_0=1/\gamma\,$ and
\begin{eqnarray}
D(p,ik)\,\approx\,D_0 \left[1+(ik)^2\, \frac {D_0}{2p} \left(
\ln^2{\frac {\gamma}{p}} +c\right) +\,...\right]\,\,\,, \label{lfd}
\end{eqnarray}
where $\,c\,$ is a numerical constant. In the dimensional notation,
evidently, we must write $\,D_0=v_0^2\tau_0/\gamma =(T/m)\tau_m\,$
and replace $\,\gamma\,$ by $\,\gamma/\tau_0\equiv 1/\tau_m\,$ in the
logarithm, with $\,\tau_m\,$ being the momentum relaxation time.

Further, we turn to direct statistical characteristics of random
diffusive (``Brownian'') displacement of the test particle as
described by the diffusion equation (\ref{de}). Knowing the first
$\,N\,$ terms of the expansion of the diffusion coefficient in
$\,ik\,$, we can find theoretically the first $\,N+2\,$ of the
statistical moments of the displacement,
\[
M_n(t)\,=\,\int R^n\,W(t,R) \,dR\,\,\,.
\]
From (\ref{de}) we have, particularly,
\[
\int_0^{\infty} M_2(t)\, e^{-\,p\,t}\, dt\,=\, \frac
{2D_0(p)}{p}\,\,\,, \,\,\,\,\, \int_0^{\infty} M_4(t)\, e^{-\,p\,t}\,
dt\,=\,\frac {24}{p^{\,2}}\left[D_1(p)+\frac
{D_0^2(p)}{p}\right]\,\,.
\]
Now, combining these formulas with (\ref{sum}) and (\ref{lfd}) and
performing the inverse Laplace transform for $\,t\gg\tau_m,\,$, we
obtain
\begin{eqnarray}
M_2(t)\,\approx\,2D_0t\,\,\,, \,\,\,\,\, M_4(t)\,\approx\,
3M_2^2(t)\,+\,6D_0^2\, t^2\left[\, \ln^2{\!\frac {t}{\tau_m}}
+c^{\prime} \ln{\frac {t}{\tau_m}} +c^{\prime\prime}\right]\,\,\,,
\label{m4}
\end{eqnarray}
where $\,D_0=T\tau_m/m\,$, and $\,c^{\prime}\,$ and
$\,c^{\prime\prime}\,$ are numerical constants.

Let us compare this result with that which would follow from the
canonic approximation based on the molecular chaos hypothesis and the
Boltzmann-Lorentz equation. At that, in place of our infinite system
of equations (\ref{fln}) we would have the single closed equation,
\[
(p+ikv_1)\,\widetilde{\phi}_1\,=\, \Lambda_1\,\widetilde{\phi}_1 +
v_1\, \widetilde{f}_{ex}\,p^{-1}A_0(v_1)\,\,\,,
\]
which at arbitrary choice of the operator $\,\Lambda\,$ implies
\begin{eqnarray}
M_2(t)\,\approx\,2D_0t\,\,\,, \,\,\,\,\, M_4(t)\,\approx\,
3M_2^2(t)\,+\,c_0\lambda_0^2 D_0 \,t\,\,\,, \label{m40}
\end{eqnarray}
with $\,c_0\,$ being a numerical constant ($\,c_0=0\,$ when
$\,\Lambda\,$ is of the EFP type) and the same diffusivity $\,D_0\,$
as above (by its sense, $\,D_0=\lim_{\,t\rightarrow\infty}
M_2(t)/2t\,$).

The second term in the expressions for $\,M_4(t)\,$ in (\ref{m4}) and
(\ref{m40}) is the fourth-order cumulant (semi-invariant) of the test
particle displacement, $\,\varkappa_4(t)\equiv M_4(t)-3M_2^2(t)\,$.
As is known, it is a measure of ``non-Gaussianity'' of the
displacement. In particular, it shows how substantially a (more or
less random) value of diffusivity $\,\widetilde{D}\,$ measured from a
concrete realization of the Brownian motion can differ from the
diffusivity $\,D_0\,$ characterizing the average over statistical
ensemble of realizations. The ``Boltzmann'' asymptotic (\ref{m40}),
that is\, $\,\varkappa_4(t)\propto t\,$\,, means that random
wandering of the test particle can be exhaustively described by the
single statistical parameter $\,D_0\,$.

This is not the case, however, if\, $\,\varkappa_4(t)\propto
t^{\,\nu}\,$\, with $\,\nu >1\,$. Such kind of the asymptotic means
that it is no longer possible to pack an arbitrary typical
realization within the framework of a single parameter $\,D_0\,$. It
is not difficult to make sure that such the asymptotic is
statistically equivalent to the existence if flicker fluctuations of
diffusivity with low-frequency power spectrum $\,\propto
\omega^{-(\,\nu -1)}\,$ \cite{bk1,bk3,pr195,bkc,bk2}.

In the theory under consideration, in contrast with the Boltzmann
theory, we are just in such the situation, since, according to
(\ref{m4}), $\,\varkappa_4(t)\propto t^2\ln^2(t/\tau_m)\,$. It can be
described, on a coarse enough time scale (when relatively small terms
proportional to $\,t\,$ in $\,\varkappa_4(t)\,$ can be neglected), as
a Gaussian random walk with a random diffusivity
$\,\widetilde{D}=\widetilde{D}(t)\,$. Treating $\,\widetilde{D}\,$ as
global characteristics of the entire time interval $\,(0,t)\,$ under
observation, we come to a ``doubly random'' walk whose path
probability density distribution and statistical moments can be
written as follow,
\begin{eqnarray}
W(t,R)\,=\, \langle\, [4\pi
t\widetilde{D}(t)]^{-1/2}\,\exp{[-R^2/4t\widetilde{D}(t)]}\,\rangle
\,\,\,, \nonumber\\
M_2(t)\,=\, \langle \,2t \widetilde{D}(t) \,\rangle\,\,\,,\,
\,\,\,\,\, M_4(t)\,=\, 3\,\langle \,[2t \widetilde{D}(t)]^2
\,\rangle\,\,\,,\, \nonumber
\end{eqnarray}
where the angle brackets mean average in respect to
$\,\widetilde{D}\,$ with $\,\langle\widetilde{D} \rangle=D_0\,$. The
corresponding interpretation of the asymptotic (\ref{m4}) in terms of
power spectrum, $\,S_D(\omega)\,$, of the diffusivity fluctuations at
$\,\omega\tau_m \ll 1\,$ yields (see Appendix B)
\begin{eqnarray}
D_0^{-2}S_D(\omega)\,\approx\,\frac {\pi}{\omega}\, \ln{\frac
{1}{\omega\tau_m}} \,\,\,. \label{sp}
\end{eqnarray}
Thus, the random diffusivity, or the ``rate of diffusion'', of the
test particle (and hence of any particle) is a random process with a
1/f spectrum. Let us discuss this result\,\footnote{\, The approach
suggested twenty years ago in the present paper
[\,Sov.\,Phys.\,JETP\, {\bf 67} (12), 2469 (1988)] later was
simplified and improved in [\,Yu.\,E.\,Kuzovlev,\, ``On statistics
and 1/f noise of Brownian motion in Boltzmann-Grad gas and finite gas
on torus. I. Infinite gas'',\, arXiv:\, cond-mat/0609515\,]. There
the asymptotic \[\,W(t,R)\approx \int_0^{\infty} \frac
{\exp{[-R^2/4\widetilde{D}t]}}{[4\pi \widetilde{D}t]^{1/2}}\,\,\,
w\left(\frac {\widetilde{D}}{D_0}\right) \frac
{d\widetilde{D}}{D_0}\,= \frac {\Gamma (5/2)}{(4\pi D_0t)^{1/2}\,
(1+R^2/4D_0t)^{\,5/2}}\,\] was found, with\, $\,w(x) =
x^{-\,3}\exp{(-\,1/x)}\,$\, representing effective probability
density distribution of the relative rate of diffusion
$\,x=\widetilde{D}/D_0\,$.}\,.

A measurement of spectrum of the diffusivity fluctuations is nothing
but a measurement of the equilibrium average value of definite
fourth-order polynomial functional of the particle velocity. An
experiment of this type is not merely thinkable but really has been
carried out by Voss and Clarke (see references in \cite{bk1,bk3}).
However, it is vastly simpler to imagine measurements of usual
quadratic functionals of the drift velocity of gas particle (under
influence of an external force) and thus spectra of fluctuations in
particle's mobility. Natural reasonings prompt that these
fluctuations should be (at least at low frequencies) a statistical
copy of the fluctuations in diffusivity. In the Appendix C we prove
that this is indeed the case for the system under our consideration
(the ``extension of the Einstein relation to fluctuations'' takes
place).

From the standpoint of a principal experimental test of the theory,
it would be interesting to extend it to multi-component gases, in
particular, weakly ionized gas. It can be shown that diffusivity
fluctuations of some component of a mixture represent a generalized
flicker noise with power spectrum $\,\omega^{-\alpha}\,$, where the
exponent takes a value from interval $\,1\leq\alpha <2\,$, depending
on ratios of masses and momentum relaxation times of the components.
Hence, the same principal mechanism of diffusivity fluctuations may
produces a variety of  flicker-type spectra. This can be illustrated
also be spectrum $\,\omega^{-1}\,\ln^{-2}{(1/\omega \tau_0)}\,$ found
in the earlier phenomenological theory (see
\cite{bk1,pr195,bk2,foot1}).

The spectrum (\ref{sp}), as well as the asymptotic (\ref{m4}), give
clear evidences that the diffusivity and mobility fluctuations behave
like statistically non-stationary and non-ergodic random processes.
This statement means that in measurements of the diffusivity or
mobility on a concrete particular realization of the particle's
random walk the variance (in the sense of the ensemble average) of
the result does not decrease, or even increases instead, when
increasing length of the observation. The probabilistic aspects of
such a behavior were studied in \cite{pr195,bk2}. Here, it is
necessary to emphasize that such the statistical non-stationarity has
no relation to a thermodynamical non-stationarity, since formulas
(\ref{lfd}) and (\ref{m4}) concern thermodynamically equilibrium gas.
The matter is that the non-stationarity manifests itself only in
dependencies on duration of observation of a test particle, but not
on time moment when the observation starts. Thus, the diffusivity and
mobility, as well as kinetic and transport coefficients in general,
can undergo a non-stationary low-frequency fluctuations even in
thermodynamically stationary and equilibrium systems. This principal
possibility should be taken in mind when interpreting experiments.

\section{Conclusion}
The 1/f\, self-diffusion noise found above is the property of an
infinitely low-density gas. Hence, the mechanism of this noise by its
very nature is indifferent to the gas density and is not related to
any dynamic many-particle correlations (e.g. through repeated
collisions). It can be related only to fluctuations in the rate and
efficiency of collisions of a given gas particle with the rest of gas
because of randomness of geometrical factors of their encounters. The
corresponding noise is just the 1/f noise if the system constantly
forgets a total number of previous collisions of the particle and
their past rates \cite{pr195,bkc,bk2}. More rigorously, the loss of
memory is implied by the property of mixing of phase trajectories of
the system in its full phase space. For a gas this property was
proved already in \cite{kr}. Notice that in fact in \cite{kr} it was
shown also that generally just the mixing of phase trajectories
causes their non-ergodic behavior, so that time-average rate of
collisions (or some other events) along a particular trajectory is
not obligated to coincide with the ensemble-average rate, regardless
of duration of the time averaging. The essential paradoxical point is
that although such the random behavior results from the loss of
memory, it can be described in the statistical language only by means
of infinitely long-living correlations.

In the derivation of kinetic equations (like the Boltzmann equation),
however, always an assumption is made (like the molecular chaos) in
order to replace actual random rate of collisions, or other
elementary kinetic events, by some ensemble-average
value\,\footnote{\, This real mission of the molecular chaos
hypothesis (Boltzmann's ``Stosszahlansatz'') still is not realized
even by important physicists. The common naive opinion is that it
claims independence of particles' velocities. But in fact it
introduces also a priori predetermined rate of collisions (although
neither the underlying mechanics nor the statistical ensemble if
initial conditions to this mechanics do present such a quantity). In
contrast to it, the ``weakened molecular chaos'' ansatz, as expressed
above in Sec.3 by the equality $\,A_{\,n+
1}^{in}(t,z,v_1\,,...,v_n|\,v_{n+ 1}) =A_1(t,z,v_{n+
1})A_{\,n}^{\prime}(t,z,v_1\,,...,v_n) \,$,\, does not dictate a
certain rate of collisions, since generally
$\,A_{\,n}^{\prime}(t,z,v_1\,,...,v_n) \neq
A_{\,n}(t,z,v_1\,,...,v_n)\,$.\,}\, \cite{pr195,bkc}. This decisive
assumption is thus introduced already into the ``zero-order'' theory
of infinitely low-density gas. Of course, taking into account finite
gas density effects (see e.g. \cite{ll2}) results in appearance of
characteristic long-scale hydrodynamic fluctuations which contribute
to both velocity and diffusivity of any particle and thus to
inter-particle correlations. But, in spite if these complications,
the above pointed out mechanism of the\, 1/f\, fluctuations anyway
remains lost\,\,\footnote{\, This fact prompts that, analogously, the
kinetic theory of electron-phonon systems in solids (in particular,
the theory of polarons) loses\, 1/f\, fluctuations in the phonon
relaxation rates and electron mobilities at very beginning of the
theory, when it attracts the hypothesis of ``random phases'' or
another ansatz to throw out statistical inter-(quasi-)particle
correlations. About that, see e.g. \cite{bk3}, the article
[\,Yu.\,Kuzovlev, ``Relaxation and 1/f noise in phonon systems'',\,
JETP\, {\bf 84}\, (6), 1138 (1997)] and preprint [Yu.\,E.\,Kuzovlev,
``Kinetic theory beyond conventional approximations and 1/f
noise'',\, arXiv:\, cond-mat/9903350\,]. Thus, in essence, the
present paper (along with \cite{bk1,bk2}) shows the way to rather
general explanation of the 1/f-noise phenomenon. }\,.

In the present work, our goal was just a correct exposure and
reconstruction of the loss, at least in the low-density limit, at
that accenting principal role of spatial non-homogeneity of gas.
Notice that the generalization of the genuine Boltzmann equation to
inhomogeneous situations by means of automatic adding of the drift
term already has attracted critical comments more than once in the
past (see e.g. \cite{kac}). The hierarchy of kinetic equations found
in Sec.3 is nothing but the correct (compatible with the concept of
collision) formulation of the Boltzmann equation for inhomogeneous
case.

At last, let us note a relation of our theory to the formally exact
generalized non-Marcovian kinetic equation for one-particle
distribution function obtained in \cite{bal} by the method of
projection operators. In \cite{bal} it was shown that analyticity of
the kernel of this equation in respect to the Laplace transform
variable ($\,p\,$ in our notation) means its coincidence in the
low-density limit with the Boltzmann equation. Hence the latter is
invalid if the kernel is nonanalytic. The asymptotic (\ref{fln})
indicates just such the case. Thus, our specific theory agrees with
the abstract theory of \cite{bal}.

I wish to thank A.\,I.\,Lomtev, G.\,N.\,Bochkov, and
Yu.\,M.\,Ivanchenko and participants of his seminar for useful
discussions.

\appendix
\section{}
The switching on at time $\,t=0\,$ a potential force field
$\,f_{ex}(r)=-\partial U(r)/\partial r\,$ acting onto the test
particle is described by the Hamiltonian
\[
H(t)\,=\, H_0 +U(q_1)\,\theta(t)\,=\,H_0+\theta(t)\sum_k \frac
{ik\,\widetilde{f}_{ex}(k)}{k^2\Omega}\,e^{ikq_1}\,\equiv\,H_0-\sum_k
X_k(t)Q_k\,\,,
\]
where $\,\theta(t)\,$ is the Heaviside step function, $\,H_0\,$ is
Hamiltonian of the unperturbed gas, and\, $\,Q_k\equiv e^{ikq_1}\,$
and\, $\,X_k(t)\equiv
-[ik\,\widetilde{f}_{ex}(k)]k^{-2}\Omega^{-1}\theta(t)\,$\, in
couples play roles of generalized variables and conjugated external
forces. Let us consider the ``flows''
\[
I_k\,\equiv\,\frac {\partial Q_k}{\partial t}\,=
ikv_1\,e^{ikq_1}\,\,\,,
\]
where $\,v_1\,$ is velocity of the text particle. According to the
Green-Kubo relations, to the first order in the forces we have
\[
\langle I_k(t)\rangle \,=\, \frac 1T \sum_{k^{\,\prime}}
\int_{-\infty}^t \langle I_k(t), I_k(t^{\,\prime})\rangle_0\,
X_{k^{\,\prime}}(t^{\,\prime})\, dt^{\,\prime}\,\,\,,
\]
where $\,\langle ...\rangle_0\,$ means the average over equilibrium
statistical ensemble of phase trajectories of the system
corresponding to the Gibbs canonic statistical ensemble of initial
states of these trajectories, and $\,\langle ...\rangle\,$ means the
non-equilibrium average. In our case, in view of the homogeneity and
isotropy of the equilibrium gas, the above relation takes the form
\begin{eqnarray}
\int \int v_1\, e^{-ikq_1}\, F_1(t,q_1,v_1)\, \frac
{dq_1}{\Omega}\,dv_1\,
\equiv\, \langle v_1\, e^{-ikq_1}\rangle \,=\, \label{a1} \\
=\,\frac 1T \int_{-\infty}^t \langle\,
v_1(t)\,\exp{\{-ik\,[q_1(t)-q_1(t^{\,\prime})] \}}\,
v_1(t^{\,\prime})\,\rangle_0\,\,dt^{\,\prime}\,\, \frac
{\widetilde{f}_{ex}(k)}{\Omega}\,\,\,. \nonumber
\end{eqnarray}
The distribution of probability density of the test particle
displacement, or path, in equilibrium gas can be determined by the
expression
\[
W(t-t_0,R)\,=\, \langle \,\delta(q_1(t)-q_1(t_0)-R)\,\rangle\,\,\,.
\]
Let us apply to it the differentiation operation
$\,\partial^2/\partial t\partial t_0\,$ and then make the Fourier
transform in respect to $\,R\,$ (at $\,t_0=0\,$) and the Laplace
transform in respect to $\,t\,$. Using $\, \langle v_1\,
\rangle_0=0\,$, we find
\begin{eqnarray}
p\,[p\,\widetilde{W}(p,ik)-1]= -\,k\int_0^{\infty} dt\, e^{-p\,t}\,
\langle\, v_1(t)\,\exp{\{-ik\,[q_1(t)-q_1(0)] \}}\,
v_1(0)\,\,\rangle_0\, k\,\,\,, \label{a2}
\end{eqnarray}
where $\,\widetilde{W}(p,ik)\,$ is the transform of
$\,\widetilde{W}(t,R)\,$. For it from (\ref{de}) we have the identity
\[
p\,\widetilde{W}(p,ik)-1\,=\,
-k\,D(p,ik)\,k\,\widetilde{W}(p,ik)\,\,\,,
\]
which, as combined with (\ref{a2}), yields
\begin{eqnarray}
\frac {D(p,ik)}{p\,+\,D(p,ik)\,k^2}\, =\,\frac 1p \int_0^{\infty}
dt\, e^{-p\,t}\, \langle\, v_1(t)\,\exp{\{-ik\,[q_1(t)-q_1(0)] \}}\,
v_1(0)\,\,\rangle_0\,\,\,.\label{a3}
\end{eqnarray}
At last, performing in (\ref{a1}) the Laplace transform and comparing
the result with (\ref{a3}), we come to the formula (\ref{gk}).

\section{}
The asymptotic of the fourth-order cumulant $\,\varkappa_4(t)\,$
expressed by (\ref{m4}) indicates non-stationary character of the
fluctuations in diffusivity (observed on a particular random walk of
the test particle). In other words, this asymptotic indicates the
absence of a certain time-average value of the diffusivity.
Therefore, $\,\widetilde{D}(t)\,$  should be treated as a
non-stationary random process (in the sense of the probability
theory). Since at $\,t\lesssim\tau_m\,$ we have $\,M_{2n}(t)\propto
t^{\,2n}\,$, this process begins from zero value:
$\,\widetilde{D}(0)=0\,$. The power spectrum of a random process of
this sort is characterized, as is known, by so-called structural
function:
\[
\langle \,[\widetilde{D}(t)-\widetilde{D}(0)]^2\,\rangle\,=\,
2\int_0^{\infty} [1-\cos{\omega t}]\,\, S_D(\omega)\,\frac
{d\omega}{2\pi}\,\,\,.
\]
At $\,t\gg\tau_m\,$, after differentiation in respect to $\,t\,$,
this expression and formulas of Sec.4 together yield:
\[
\int_0^{\infty} \omega S_D(\omega)\, \sin{\omega t}\,\,\frac
{d\omega}{\pi}\,=\, \frac {d}{dt}\, \frac {M_4(t)}{12\,t^2}\,
\approx\, \frac {D_0^2}{t}\, \ln{\frac {t}{\tau_m}}\,\,\,.
\]
The spectrum (\ref{sp}) follows immediately.

\section{}
Let at time $\,t=0\,$ a constant force $\,f\,$ begins to act upon the
test particle (``constant'' means that $\,f\,$ is independent on
$\,t\,$ and $\,q_1\,$), and $\,W(t,R;f)\,$ denotes resulting
probability distribution of the particle's displacement,
$\,R(t)=q_1(t)-q_1(0)\,$. According to the generalized
fluctuation-dissipation relations (see e.g. \cite{fds} and
corresponding use of these relations in \cite{bk3}), we can
write\,\footnote{\, Besides \cite{fds}, the generalized
fluctuation-dissipation relations were investigated e.g. in the works
[G.\,N.\,Bochkov and Yu.\,E.\,Kuzovlev,\, ``Fluctuation-dissipation
relations for nonequilibrium processes in open systems'',\,
Sov.\,Phys.\,JETP\, {\bf 49},\, 543\, (1979);\,
``Fluctuation-dissipation theory of nonlinear viscosity'',\,
Sov.\,Phys.\,JETP\, {\bf 52}.\, No.12 (1980); ``Nonlinear
fluctuation-dissipation relations and stochastic models in
nonequilibrium thermodynamics. I. Generalized fluctuation-dissipation
theorem'',\, Physica {\bf A 106},\, 443-480 (1981)].}\,
\begin{equation}
W(t,R;f)\,\exp{(-fR/T)}\,=\,W(t,-R;f)\,\,\,. \label{a4}
\end{equation}
It is not difficult to transform this exact equality into
\begin{equation}
W(t,R;f)-W(t,-R;f)\,=\,\tanh{\left(\frac {fR}{2T}\right)}\,
\,[\,W(t,R;f)+W(t,-R;f)\,]\,\,\,. \label{a5}
\end{equation}
Multiplying this by $\,R\,$ and integrating over $\,R\,$, we find
such a consequence of (\ref{a4}):
\begin{equation}
\langle\,R(t)\,\rangle\,=\, \left\langle\,R(t)\, \tanh{\frac
{fR(t)}{2T}}\, \right\rangle\,\,\,. \label{a6}
\end{equation}

Now, consider the third derivative of this equality with respect to
$\,f\,$ at $\,f=0\,$:
\begin{equation}
\left[\frac {\partial^{\,3} \langle R(t)\rangle}{\partial
f^3}\right]_{f=0}\,=\, \frac {3}{2T}\,\left[\frac {\partial^{\,2}
\langle R^{\,2}(t)\rangle }{\partial f^2}\right]_{f=0} -\,\frac
{\langle \,R^{\,4}(t)\,\rangle_0}{4T^3}\, \,\,\,, \label{a7}
\end{equation}
and apply this general relation to our problem. The average
displacement
\[
\langle \,R(t)\,\rangle\,=\, \int_0^t \langle\,
v_1(t^{\,\prime})\,\rangle\,\, dt^{\,\prime}\,
\]
under the influence of the constant force $\,f_{ex}=f=\,$const\, is
determined by the homogeneous solution to equations (\ref{ln}). In
the homogeneous case, however, they evidently reduce to the ordinary
Boltzmann-Lorentz equation. We can thus assert that the average drift
velocity, $\,\langle\, v_1(t)\,\rangle\,$, reaches saturation at
$\,t\gtrsim \tau_m\,$. Consequently, the left side of (\ref{a7})
increases linearly with the time. However, the last term in
(\ref{a7}), which contains the fourth-order statistical moment of
equilibrium displacement, $\,\langle \,R^{\,4}(t)\,\rangle_0
=M_4(t)\,$, due to (\ref{m4}) increases far more rapidly. Hence, this
term is compensated (exactly at $\,t/\tau_m\rightarrow\infty\,$) by
the first right-hand term of (\ref{a7}). As the result, from
(\ref{a6}) and (\ref{a7}) we obtain
\begin{eqnarray}
\langle \,R(t)\,\rangle\,=\,\frac {f}{2T}\,M_2(t)\, \approx\, \frac
{D_0}{T}\,ft\,\,\,,\nonumber \\
\langle \,R^2(t)\,\rangle\,=\, M_2(t)+\frac
{f^2}{12T^2}\,M_4(t)\,\approx \,2D_0t + \frac {f^2t^2}{T^2}\, \langle
\,\widetilde{D}^2(t)\,\rangle\,\,\,. \label{a8}
\end{eqnarray}

On the other hand, we can write
\[
\langle \,R(t)\,\rangle\,=\,t\,\langle \,\mu(t)\,\rangle\,f\,\,,
\,\,\,\,\,\,\, \langle \,R^2(t)\,\rangle\,=\,M_2(t)\,+\,t^2\,\langle
\,\mu^2(t)\,\rangle\,f^2\,\,\,,
\]
where\, $\,\mu(t)\,$ is the mobility referred to the entire
observation interval as a whole. A comparison with (\ref{a8}) shows
that the Einstein relation between the diffusivity and the mobility
holds not only in the ordinary sense (for the ensemble-average
values) but also for their fluctuations. Correspondingly, the
spectrum (\ref{sp}) simultaneously refers to the relative
fluctuations in mobility. A similar relation, concerning the spectral
density of equilibrium electric voltage noise and electrical
conductivity, has been confirmed experimentally in the famous
Voss-Clarke experiment (see e.g. \cite{bk3}).

\end{document}